\documentclass[aip,jcp,reprint,amsmath,amssymb,floatfix]{revtex4-2}

\usepackage{bm}
\usepackage{graphicx}
\graphicspath{{figure/}{./}}
\usepackage{xcolor}
\usepackage[breaklinks=true]{hyperref}
\hypersetup{
  colorlinks = true,
  linkcolor  = {blue!55!black},
  citecolor  = {green!45!black},
  urlcolor   = {blue!70!black},
  linktoc    = all,
}

\newcommand{\Lph}{L_{\mathrm{ph}}}
\newcommand{\avg}[1]{\langle #1 \rangle}
\newcommand{\sgn}{\operatorname{sgn}}
\newcommand{\PH}{\mathrm{PH}}
\newcommand{\SP}{\mathrm{SP}}
\newcommand{\gsr}{g_{\mathrm{sr}}}
\newcommand{\Hsr}{H_{\mathrm{SR}}}
\newcommand{\Gcap}{\Gamma_{\mathrm{CAP}}}

\begin{document}

\title[Additive control of CISS by phonon chirality]{Phonon chirality as an additive control of CISS: a symmetry-protected law}

\author{Shi-Qi Zhang}
\affiliation{Key Laboratory for Quantum Materials of Zhejiang Province,
Department of Physics, School of Science and Research Center for Industries of
the Future, Westlake University, Hangzhou, Zhejiang 310030, China}

\author{Vipul Upadhyay}
\affiliation{Department of Chemistry, Bar-Ilan University, Ramat-Gan 52900, Israel}
\affiliation{Institute of Nanotechnology and Advanced Materials, Bar-Ilan University,
Ramat-Gan 52900, Israel}
\affiliation{Center for Quantum Entanglement Science and Technology, Bar-Ilan University,
Ramat-Gan 52900, Israel}

\author{Jiayue Han}
\affiliation{Department of Chemistry, School of Science and Research Center for
Industries of the Future, Westlake University, Hangzhou, Zhejiang 310030, China}
\affiliation{Institute of Natural Sciences, Westlake Institute for Advanced
Study, Hangzhou, Zhejiang 310024, China}

\author{Amikam Levy}
\affiliation{Department of Chemistry, Bar-Ilan University, Ramat-Gan 52900, Israel}
\affiliation{Institute of Nanotechnology and Advanced Materials, Bar-Ilan University,
Ramat-Gan 52900, Israel}
\affiliation{Center for Quantum Entanglement Science and Technology, Bar-Ilan University,
Ramat-Gan 52900, Israel}

\author{Wenjie Dou}
\email{douwenjie@westlake.edu.cn}
\affiliation{Department of Chemistry, School of Science and Research Center for
Industries of the Future, Westlake University, Hangzhou, Zhejiang 310030, China}
\affiliation{Institute of Natural Sciences, Westlake Institute for Advanced
Study, Hangzhou, Zhejiang 310024, China}
\affiliation{Key Laboratory for Quantum Materials of Zhejiang Province,
Department of Physics, School of Science and Research Center for Industries of
the Future, Westlake University, Hangzhou, Zhejiang 310030, China}

\date{\today}

\begin{abstract}
Chirality-induced spin selectivity (CISS) is usually associated with molecular
handedness. The possible contribution of chiral phonons is less established.
We study a helical tight-binding model in which local phonon angular momentum
modulates spin-dependent nearest-neighbor hopping. Fewest-switches surface
hopping calculations give the transmitted spin polarization
$\SP=aC+b\,\PH$. Here $C$ is the molecular chirality and $\PH$ is the phonon
chirality. A mirror symmetry reverses $C$, $\PH$, and $\SP$ simultaneously.
This symmetry excludes both a chirality-independent offset and a $C\cdot\PH$ term.
The phonon contribution can therefore enhance, cancel, or reverse the molecular
CISS signal.
\end{abstract}

\maketitle

\section{Introduction}
\label{sec:intro}

Chirality-induced spin selectivity (CISS) encompasses electronic responses
that correlate molecular handedness with electron spin
~\cite{NaamanReview,NaamanRevChem2019}. Experiments probe these responses
through spin-resolved electron transmission and photoemission, as well as
magnetoresistance measurements in junctions containing chiral molecules
~\cite{Ray1999,Gohler2011,Xie2011,Mishra2013,Kettner2018,Kondou2022}.
In the transport setup considered here, we quantify CISS by the spin
polarization of electrons collected at one end of a chiral molecule after
unpolarized injection at the other end. This observable describes the present
setup rather than all experimental manifestations of CISS. A broadly accepted
microscopic explanation of the large room-temperature signals and their
robustness across different experimental settings remains lacking
~\cite{EversRoadmap,Bloom2024}.

Explaining spin-selective transmission must account for the constraints
imposed by time-reversal symmetry. Consider a nonmagnetic molecule connected
to two nonmagnetic leads, each supporting one orbital channel with two spin
states, with coherent and elastic transport. Time-reversal symmetry then
requires equal transmission eigenvalues, so an unpolarized incident current
remains unpolarized after transmission~\cite{Bardarson2008,Aharony2025}.
Additional orbital channels can permit spin-polarized transmission while
preserving time-reversal symmetry~\cite{Utsumi2020,Utsumi2022}, whereas
leakage to the environment changes the unitary two-terminal scattering
problem~\cite{Matityahu2016}. Electron--phonon and electron--electron
interactions provide further routes beyond the coherent, elastic
single-channel description~\cite{FranssonPRB2020,Fransson2021}.

Theoretical proposals also differ in where spin selectivity originates.
Spinterface models emphasize the interplay of molecular electron motion and
spin-dependent processes at molecule--electrode interfaces
~\cite{AlwanDubi2021,Dubi2022}. For donor--chiral-bridge--acceptor systems,
Fay and Limmer proposed spin polarization generated by spin--orbit and
exchange couplings during incoherent electron transfer~\cite{FayLimmer2021}.
Chiesa \textit{et al.} studied many-body correlations on the chiral bridge
and the role of vibrations in the long-time acceptor polarization
~\cite{Chiesa2024}. Other proposals for this geometry involve dephasing-assisted
spin dynamics~\cite{Zhang2025}, vibrationally mediated
Dzyaloshinskii--Moriya interactions~\cite{Chiesa2026}, and a spinterface-like
interaction within the donor--bridge--acceptor complex~\cite{Sarkar2026}.
These studies motivate a distinction between polarization of a molecular
charge-separated state and spin polarization of a collected electron flux.

Nuclear motion can affect electron transport by changing the electronic
Hamiltonian. In Fransson's helical tight-binding model, nuclear displacements
modulate electronic hopping and spin--orbit coupling~\cite{FranssonPRB2020}.
Mixed quantum--classical studies have examined correlations between
nonequilibrium vibrations and spin selectivity~\cite{Smorka2025}, while
calculations for a related molecular junction found transient polarization
that decayed nearly to zero at long times~\cite{HanWangDou}.
Self-consistent nonequilibrium Green's function (NEGF) calculations found
negligible polarization in the weak electron--phonon coupling regime over the
investigated parameter range~\cite{UpadhyayLevy2026}. In contrast, a fully
quantum treatment of vibronic dynamics found substantial current polarization
in an off-resonant, low-voltage regime~\cite{Rudge2025}. The role of vibrations
therefore depends on the coupling, transport regime, and treatment of the
vibrational dynamics; their inclusion alone does not guarantee a large
polarization.

An additional possibility is that phonon angular momentum can influence
spin-selective transport. Two mutually orthogonal transverse vibrations have
circular combinations that carry opposite angular momenta
~\cite{ZhangNiu2014,Zhu2018}. Studies of phonon-induced spin polarization and
spin transport have explored how this rotational motion couples to electron
spin~\cite{FranssonPRR2023,Qin2025,Kim2023}. Related proposals include
phonon--spin conversion at an interface~\cite{Funato2024} and spin-vorticity
coupling during phonon-assisted variable-range hopping along DNA
~\cite{SanoKato2024}. Nuomin \textit{et al.} showed that directional electronic
current can generate phonon angular momentum with the chirality and
current-reversal properties associated with CISS~\cite{Nuomin2026}.
Here we instead prescribe a circular-mode bias and ask how phonon chirality,
controlled independently of molecular handedness, modifies the collected
spin polarization.

To address these questions, we introduce a coupling between local phonon angular
momentum and spin-dependent hopping into Fransson's model. The label $C$
specifies molecular chirality, whereas $\PH$ specifies the sign of the
frequency splitting between the two circular vibrational modes. We propagate
the coupled electron--phonon dynamics with an FSSH/mean-field scheme based on
fewest-switches surface hopping (FSSH)~\cite{Tully1990}. A spin-independent
complex absorbing potential (CAP) collects electronic population at the
terminal site, allowing us to determine the spin polarization of the collected
electrons. We find that the spin selectivity obeys the following
additive form
\[
  \SP=aC+b\,\PH.
\]
The numerical results support this symmetry-derived relation, which identifies
phonon chirality as an additional control variable for spin polarization.

Section~\ref{sec:model} introduces the model, and Sec.~\ref{sec:method}
describes the dynamics. Section~\ref{sec:additive} derives the additive law.
The results, discussion, and conclusions are presented in
Secs.~\ref{sec:results}, \ref{sec:discussion}, and \ref{sec:conclusions},
respectively.

\section{Model Hamiltonian}
\label{sec:model}

We extend the helical tight-binding model of Ref.~\cite{FranssonPRB2020}
with two transverse vibrational coordinates at each site and a coupling
between their angular momentum and spin-dependent hopping. The total
Hamiltonian is
\begin{equation}
\begin{aligned}
  H_{\mathrm{tot}}(\bm R,\bm P)={}&H_{\mathrm{el}}
  +H_{\mathrm{el-ph}}(\bm R)\\
  &+H_{\mathrm{vib}}(\bm R,\bm P)
  +\Hsr(\bm R,\bm P).
\end{aligned}
\label{eq:Htotal}
\end{equation}

The static electronic Hamiltonian is
\begin{equation}
\begin{aligned}
  H_{\mathrm{el}}={}&
  -t_0\sum_{j=1}^{N-1}\sum_{s=\pm1}
  \left(c^\dagger_{js}c_{j+1,s}+\mathrm{h.c.}\right)\\
  &+i\lambda_0\sum_{j=1}^{N-2}\sum_{s,s'=\pm1}
  \Bigl[(\bm v_j\!\cdot\!\bm\sigma)_{ss'}
  c^\dagger_{js}c_{j+2,s'}-\mathrm{h.c.}\Bigr].
\end{aligned}
\label{eq:Hel}
\end{equation}
Here $j=1,\ldots,N$ labels the molecular sites, and $s,s'=\pm1$ label the
two spin states: $+1$ for $\uparrow$ and $-1$ for $\downarrow$. The spin
quantization axis is the helix axis, denoted by $z$. The operators
$c^\dagger_{js}$ and $c_{js}$ create and annihilate an electron in state
$(j,s)$, and $\bm\sigma=(\sigma^x,\sigma^y,\sigma^z)$ denotes the Pauli
matrices. The parameters $t_0$ and $\lambda_0$ are the spin-conserving
nearest-neighbor hopping and static next-nearest-neighbor spin--orbit coupling
(SOC) strengths.

The SOC vector $\bm v_j$ is determined by the equilibrium geometry. We place
the molecular sites on the helix
\begin{equation}
\begin{aligned}
  \bm R_j^{(0)}&=
  \left(r_{\mathrm h}\cos\phi_j,\,r_{\mathrm h}\sin\phi_j,\,h_j\right),\\
  \phi_j&=C\,\frac{2\pi N_{\mathrm{lap}}}{N-1}(j-1),\\
  h_j&=h_{\mathrm{tot}}\frac{j-1}{N-1},
  \qquad j=1,\ldots,N.
\end{aligned}
\label{eq:helix}
\end{equation}
Here $r_{\mathrm h}$ is the helix radius, $h_{\mathrm{tot}}$ is the total
axial height from the first to the last site, and $N_{\mathrm{lap}}$ is the
number of turns; the axial pitch per turn is therefore
$p=h_{\mathrm{tot}}/N_{\mathrm{lap}}$. The labels $C=\pm1$ specify the two
enantiomers by reversing the azimuthal winding while leaving the axial
coordinates unchanged. In the dimensionless geometry used in the calculations,
$r_{\mathrm h}=1$, $h_{\mathrm{tot}}=1$, and $N_{\mathrm{lap}}=3$. These
positions define the bond directions and the SOC vector
\begin{equation}
  \bm v_j=\hat{\bm d}_{j,j+1}\times\hat{\bm d}_{j,j+2},
  \qquad
  \hat{\bm d}_{j,k}=
  \frac{\bm R_j^{(0)}-\bm R_k^{(0)}}
  {|\bm R_j^{(0)}-\bm R_k^{(0)}|}.
\label{eq:socvector}
\end{equation}
The unit vector $\hat{\bm d}_{j,k}$ points from site $k$ to site $j$.
Reversing $C$ changes the helix and hence $\bm v_j$, through which molecular
chirality enters $H_{\mathrm{el}}$. In the straight-chain limit, the bond
directions are collinear and $\bm v_j=0$.

\subsection{Electron--phonon coupling Hamiltonian}
\label{sec:elph}

Nuclear displacements modify the hopping and SOC of the static model. We
describe their leading effect by couplings linear in the transverse relative
displacements:
\begin{equation}
\begin{aligned}
  H_{\mathrm{el-ph}}&=H_{t_1}+H_{\lambda_1},\\
  H_{t_1}&=-\frac{t_1}{\sqrt2}
  \sum_{j=1}^{N-1}Q_j\sum_{s=\pm1}\hat T_{j,s},\\
  H_{\lambda_1}&=\frac{i\lambda_1}{\sqrt2}
  \sum_{j=1}^{N-2}\sum_{s,s'=\pm1}
  \Bigl[(\bm W_j\!\cdot\!\bm\sigma)_{ss'}
  c^\dagger_{js}c_{j+2,s'}-\mathrm{h.c.}\Bigr].
\end{aligned}
\label{eq:Helph}
\end{equation}
Here
\begin{equation}
  \hat T_{j,s}=c^\dagger_{js}c_{j+1,s}
  +c^\dagger_{j+1,s}c_{js}
\label{eq:Tjs}
\end{equation}
is the spin-conserving nearest-neighbor bond operator. To express the
geometric factors $Q_j$ and $\bm W_j$, we construct a local discrete Frenet
frame $(\hat{\bm T}_j,\hat{\bm N}_j,\hat{\bm B}_j)$ from the equilibrium helix
geometry~\cite{HuDiscreteFrenet2011}. The two transverse vibrational coordinates
are taken along the local normal and binormal directions,
\begin{equation}
  \bm u_j=R_{j,1}\hat{\bm N}_j+R_{j,2}\hat{\bm B}_j,
  \label{eq:displacement}
\end{equation}
where $R_{j,1}$ and $R_{j,2}$ describe motion along the normal and binormal,
respectively. With $\hat{\bm d}_j\equiv\hat{\bm d}_{j,j+1}$ and
$\hat{\bm d}^{(2)}_j\equiv\hat{\bm d}_{j,j+2}$, the relative displacements
give
\begin{align}
  Q_j&=(\bm u_{j+1}-\bm u_j)\!\cdot\!\hat{\bm d}_j,
  \label{eq:Qj}\\
  \bm W_j&=\hat{\bm d}^{(2)}_j\times(\bm u_j-\bm u_{j+2}).
  \label{eq:Wj}
\end{align}
Thus, $H_{t_1}$ modulates spin-conserving hopping through the relative
displacement along a nearest-neighbor bond, while $H_{\lambda_1}$ modulates
next-nearest-neighbor SOC. The parameters $t_1$ and $\lambda_1$ set the
respective coupling strengths.

\subsection{Vibrational Hamiltonian}
\label{sec:chiralphonon}

The vibrational Hamiltonian for the same transverse coordinates is
\begin{equation}
\begin{aligned}
  H_{\mathrm{vib}}&=H_0+H_{\Omega}+H_{\parallel},\\
  H_0&=\frac{\hbar\omega_0}{2}
  \sum_{j=1}^{N}\sum_{\alpha=1}^{2}
  \left(R_{j,\alpha}^2+P_{j,\alpha}^2\right),\\
  H_{\Omega}&=\hbar\Omega\sum_{j=1}^{N}L_{\mathrm{ph},j},
  \qquad
  L_{\mathrm{ph},j}=R_{j,1}P_{j,2}-R_{j,2}P_{j,1},\\
  H_{\parallel}&=\frac{K_{\parallel}}{2}
  \sum_{j=1}^{N-1}\sum_{\alpha=1}^{2}
  \left(R_{j+1,\alpha}-R_{j,\alpha}\right)^2.
\end{aligned}
\label{eq:Hvibsplit}
\end{equation}
Here $P_{j,\alpha}$ is the dimensionless momentum conjugate to
$R_{j,\alpha}$, with $\alpha=1,2$, and $\hbar L_{\mathrm{ph},j}$ is the local
phonon angular momentum. The term $H_0$ describes two degenerate transverse
oscillators. The angular-momentum term $H_{\Omega}$ splits their circular
combinations, while $H_{\parallel}$ couples matching transverse components
on adjacent sites.

To control the relative frequencies of the two circular modes independently
of molecular chirality, we introduce the binary label $\PH$ for the sign of
the circular-mode splitting. For $K_{\parallel}=0$, the mode frequencies and
splitting are
\begin{equation}
\begin{aligned}
  \omega_\pm&=\omega_0\pm\Omega,\\
  \Omega&=\PH\,c_f\omega_0\sin\!\left(\frac{\Delta\phi}{2}\right),\\
  \Delta\phi&=\frac{2\pi N_{\mathrm{lap}}}{N-1},
  \qquad \PH=\pm1.
\end{aligned}
\label{eq:Omega}
\end{equation}
The parameter $c_f\ge0$ controls the splitting magnitude, and stability
requires $|\Omega|<\omega_0$. For the geometry used here, $c_f>0$ gives
$\PH=\sgn\Omega$, whereas $c_f=0$ leaves the modes degenerate and $\PH$
has no dynamical effect. Because the two circular modes carry opposite angular
momenta, the splitting changes their thermal occupations. For the uncoupled
thermal distribution at $T>0$ and $\Omega\ne0$, the present mode convention
gives $\sgn\avg{\Lph}=-\PH$ (Appendix~\ref{app:modemixing}). Thus,
$\PH$ labels the sign of the circular-mode splitting rather than the sign of
the resulting mean phonon angular momentum. A fixed nonzero $\Omega$ breaks
time-reversal symmetry in the phonon sector because $L_{\mathrm{ph},j}$
changes sign under time reversal. In this model, $\Omega$ is an imposed
angular-momentum bias, not a splitting derived from nonmagnetic molecular
geometry alone. The geometric factor in Eq.~\eqref{eq:Omega} only
parametrizes its magnitude. We do not model the external preparation or
driving that would establish this bias.

\subsection{Spin--phonon coupling Hamiltonian}
\label{sec:hsr}

To describe the coupling of phonon angular momentum to electron spin, we
introduce the effective spin--phonon coupling
\begin{equation}
\begin{aligned}
  \Hsr&=-\gsr\sum_{j=1}^{N-1}
  \bar L_{\mathrm{ph},j}\,\hat B_j^z,\\
  \bar L_{\mathrm{ph},j}&=
  \frac{L_{\mathrm{ph},j}+L_{\mathrm{ph},j+1}}{2},
  \qquad
  \hat B_j^z=\sum_{s=\pm1}s\,\hat T_{j,s}.
\end{aligned}
\label{eq:HSR}
\end{equation}
Here $\bar L_{\mathrm{ph},j}$ is the bond-averaged phonon angular momentum,
and $\hat B_j^z$ is the spin-weighted bond operator. Combining $\Hsr$ with the
static nearest-neighbor hopping shows how this coupling acts on the electrons:
\begin{equation}
  H_{t_0}+\Hsr=-\sum_{j=1}^{N-1}\sum_{s=\pm1}
  \left(t_0+s\gsr\bar L_{\mathrm{ph},j}\right)\hat T_{j,s}.
  \label{eq:combinedhop}
\end{equation}
Phonon angular momentum therefore shifts the spin-up and spin-down hopping
amplitudes by equal magnitudes and opposite signs. The term $\Hsr$ is an
effective coupling introduced at the model level rather than a microscopic
derivation from the underlying molecular Hamiltonian. The parameter $\gsr$
has units of energy and sets its strength. Unlike $H_{\Omega}$ at fixed
$\Omega$, $\Hsr$ is invariant under time reversal,
because both $\bar L_{\mathrm{ph},j}$ and $\hat B_j^z$ change sign. The coupling
is also Hermitian. This symmetry of the interaction must be distinguished
from the symmetry of the phonon state. Replacing $\bar L_{\mathrm{ph},j}$ by a
prescribed mean $\ell_j=\langle\bar L_{\mathrm{ph},j}\rangle$ gives the
electronic mean-field term
$H_{\mathrm{SR}}^{\mathrm{MF}}=-\gsr\sum_j\ell_j\hat B_j^z$.
For a fixed nonzero $\ell_j$, electronic time reversal changes the sign of
this term: invariance requires reversing the prescribed phonon angular
momentum as well. Thus a time-reversal-invariant spin--phonon interaction
can produce a time-reversal-breaking electronic mean field when the phonon
state has a prescribed nonzero mean angular momentum.

\section{Dynamical method}
\label{sec:method}

\subsection{Complex absorbing potential}
\label{sec:cap}

To measure electron transfer along the chain, we place a spin-independent
complex absorbing potential (CAP) on the terminal site. In atomic units
($\hbar=1$), the CAP matrix and electronic evolution equation are
\begin{equation}
\begin{aligned}
  \Gcap&=\gamma_{\mathrm{CAP}}
  \sum_{s=\pm1}|N,s\rangle\langle N,s|,\\
  \dot\rho&=-i[H_{\mathrm e},\rho]-\frac12\{\Gcap,\rho\}.
\end{aligned}
\label{eq:lindblad}
\end{equation}
Here $H_{\mathrm e}=H_{\mathrm{el}}+H_{\mathrm{el-ph}}+\Hsr$. The decrease
in $\operatorname{Tr}\rho$ equals the population collected by the
CAP~\cite{CAPselfenergy,CAPmoljunction,CAPspintransport}. Integrating the
spin-resolved loss up to the final propagation time $t_f$ gives the collected
populations and their polarization:
\begin{align}
  J_s&=\int_0^{t_f}\!dt\,\gamma_{\mathrm{CAP}}\rho_{Ns,Ns}(t),
  \label{eq:Jsintegrated}\\
  \SP&=\frac{J_{\uparrow}-J_{\downarrow}}
  {J_{\uparrow}+J_{\downarrow}}.
  \label{eq:SPdef}
\end{align}
The observable $\SP$ therefore describes the electrons collected after
initialization, rather than a steady-state two-terminal current.

\subsection{Surface-hopping dynamics}
\label{sec:dynamics}

The collected polarization depends on the coupled electron--phonon
dynamics. We propagate $\rho$ quantum mechanically and $(\bm R,\bm P)$
classically using a fewest-switches surface-hopping (FSSH) scheme with a
mean-field treatment of $\Hsr$~\cite{Tully1990}. The position-dependent
Hamiltonian
\begin{equation}
  H_{\mathrm{SH}}(\bm R)=H_{\mathrm{el}}+H_{\mathrm{el-ph}}(\bm R)
  \label{eq:HSH}
\end{equation}
defines the adiabatic surfaces, derivative couplings, and active-surface
forces. Exactly degenerate states are treated as a single hopping subspace,
as detailed in Supplementary Note~1. The electronic density matrix is propagated with
$H_{\mathrm e}=H_{\mathrm{SH}}+\Hsr$.
Because the CAP reduces $\operatorname{Tr}\rho$, the mean-field contribution
from $\Hsr$ is evaluated with the normalized surviving density matrix
\begin{equation}
  \widetilde\rho=\frac{\rho}{\operatorname{Tr}\rho}.
\end{equation}
Since $\Hsr$ depends on the nuclear momenta, its mean-field expectation also
modifies the nuclear velocity,
\begin{equation}
  \dot{\bm R}=\dot{\bm R}_{\mathrm{bare}}
  +\frac{\partial\langle\Hsr\rangle_{\widetilde\rho}}{\partial\bm P},
  \qquad
  \langle\Hsr\rangle_{\widetilde\rho}
  =\operatorname{Tr}(\widetilde\rho\Hsr).
\end{equation}
The resulting velocity is used in the usual nonadiabatic FSSH contribution
$\dot{\bm R}\!\cdot\!\bm d_{ab}$, so $\Hsr$ affects this term indirectly through
$\dot{\bm R}$. In addition, the off-diagonal matrix elements of $\Hsr$ in the
$H_{\mathrm{SH}}$ adiabatic basis are included directly in the hopping rate.
A full phase-space surface-hopping treatment can be used for electronic
Hamiltonians that depend on both nuclear coordinates and momenta
\cite{BianPSSH2024}; such a treatment is not used here because of the size of
the present electronic Hilbert space. The real momentum-rescaling direction
is obtained by a phase selection of the complex coupling vector. No explicit
Berry-curvature force is added to the nuclear equations of motion; the
rescaling prescription and this approximation are specified in Supplementary
Note~1~\cite{MiaoSubotnik2019}.

The initial phonon coordinates and momenta are sampled from the thermal Wigner
distributions of the circular modes for the uncoupled local vibrations
($K_{\parallel}=0$ in Eq.~\eqref{eq:Hvibsplit}). To represent unpolarized
injection, half of the electronic trajectories start in the pure state
$|1,\uparrow\rangle$ and half in $|1,\downarrow\rangle$. Supplementary
Note~1 gives the equations of motion and numerical propagation scheme.

\section{The additive law \texorpdfstring{$\SP=aC+b\,\PH$}{SP = a C + b PH}}
\label{sec:additive}

Before presenting our results, we first derive an additive law based purely on symmetry analysis. 
To separate the dependence on molecular and phonon chirality, we compare
$\SP$ for all four combinations of $C,\PH\in\{+1,-1\}$ at fixed values of
the remaining parameters. These four values have the unique expansion
\begin{equation}
  \SP(C,\PH)=\alpha+aC+b\,\PH+c\,C\cdot\PH.
  \label{eq:general4sign}
\end{equation}
Writing $\SP_{++}$, $\SP_{+-}$, $\SP_{-+}$, and $\SP_{--}$ for the four
sign combinations, the coefficients follow directly as
\begin{equation}
\begin{aligned}
  \alpha&=\tfrac14
  (\SP_{++}+\SP_{+-}+\SP_{-+}+\SP_{--}),\\
  a&=\tfrac14
  (\SP_{++}+\SP_{+-}-\SP_{-+}-\SP_{--}),\\
  b&=\tfrac14
  (\SP_{++}-\SP_{+-}+\SP_{-+}-\SP_{--}),\\
  c&=\tfrac14
  (\SP_{++}-\SP_{+-}-\SP_{-+}+\SP_{--}).
\end{aligned}
\label{eq:fourCoefficients}
\end{equation}

The mirror transformation restricts which terms in this expansion can
remain. Reflection through a plane containing the transport axis maps
$(C,\PH)$ to $(-C,-\PH)$ and reverses the measured axial spin component.
The spin-independent CAP and the unpolarized electronic initial ensemble are
unchanged. When the phonon preparation and propagation respect the same
mapping, the collected polarization satisfies
\begin{equation}
  \SP(-C,-\PH)=-\SP(C,\PH).
  \label{eq:mirrorMain}
\end{equation}
Since the terms $\alpha$ and $cC\cdot\PH$ are even
under the joint reversal, Eq.~\eqref{eq:mirrorMain} requires
$\alpha=c=0$, leaving
\begin{equation}
  \SP(C,\PH)=aC+b\,\PH.
  \label{eq:additivelaw}
\end{equation}
The additive form therefore follows from the joint mirror relation, whereas
the coupled dynamics determine $a$ and $b$. The relation is exact for the two
binary labels and does not require linear dependence on continuous parameters
such as $\gsr$ or $\Omega$. 
\section{Results}
\label{sec:results}

Unless stated otherwise, the calculations use
$N_{\mathrm{lap}}=3$, $N=15$, $r_{\mathrm h}=h_{\mathrm{tot}}=1$,
$t_0=25$~meV, $t_1=8$~meV, $\lambda_0=10$~meV,
$\lambda_1=10$~meV, $\hbar\omega_0=5$~meV,
$k_BT=5$~meV, $K_{\parallel}=0$, and
$\gamma_{\mathrm{CAP}}=25$~meV. Each trajectory is propagated to
$15000$~fs with a time step of $\Delta t=10$~a.u. ($\approx0.242$~fs).
Each reported result is obtained from a total of $10^4$ electronic
trajectories. Error bars denote one standard deviation across independent
ensemble calculations.

\begin{figure}
  \centering
  \includegraphics[width=\linewidth]{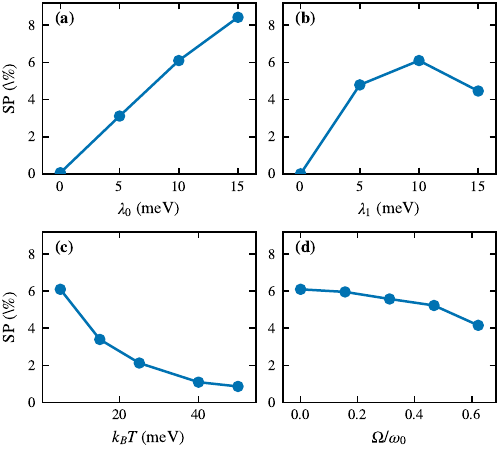}
  \caption{CAP-collected spin polarization at $\gsr=0$ as a function of
    (a)~the static SOC strength $\lambda_0$,
    (b)~the displacement-induced SOC strength $\lambda_1$,
    (c)~the vibrational sampling temperature $k_BT$, and
    (d)~the circular-mode splitting $\Omega/\omega_0$.
    Error bars denote one standard deviation across independent ensemble
    estimates.
    \label{fig:scans}}
\end{figure}

\subsection{Spin polarization without \texorpdfstring{$H_{\mathrm{SR}}$}{H(SR)}}
\label{sec:molecular-soc}

We first set $\gsr=0$ to examine spin polarization in the absence of
$\Hsr$. For the baseline parameters, independent calculations give
$\SP=0.0610\pm0.0009$ (Fig.~S1).

Figure~\ref{fig:scans}(a) shows the dependence of $\SP$ on the static SOC
strength $\lambda_0$. As $\lambda_0$ increases from 0 to 15~meV, $\SP$
increases approximately linearly from nearly zero to about $0.084$.

Figure~\ref{fig:scans}(b) shows the dependence of $\SP$ on the
displacement-induced SOC strength $\lambda_1$. At $\lambda_1=0$, $\SP$ is
nearly zero. As $\lambda_1$ increases, $\SP$ increases and reaches the
largest value among the calculated points near $\lambda_1=10$~meV, before
decreasing at $\lambda_1=15$~meV. Thus, setting either $\lambda_0=0$ or
$\lambda_1=0$ gives a spin polarization close to zero.

Figure~\ref{fig:scans}(c) shows the effect of the initial vibrational sampling
temperature. Increasing $k_BT$ from 5 to 50~meV reduces $\SP$ from about
$0.061$ to about $0.009$. By contrast, varying the frequency splitting
$\Omega/\omega_0$ between the two circularly polarized vibrational modes at
$\gsr=0$ produces only a small change in $\SP$ [Fig.~\ref{fig:scans}(d)].

\begin{figure*}
  \centering
  \includegraphics[width=\textwidth]{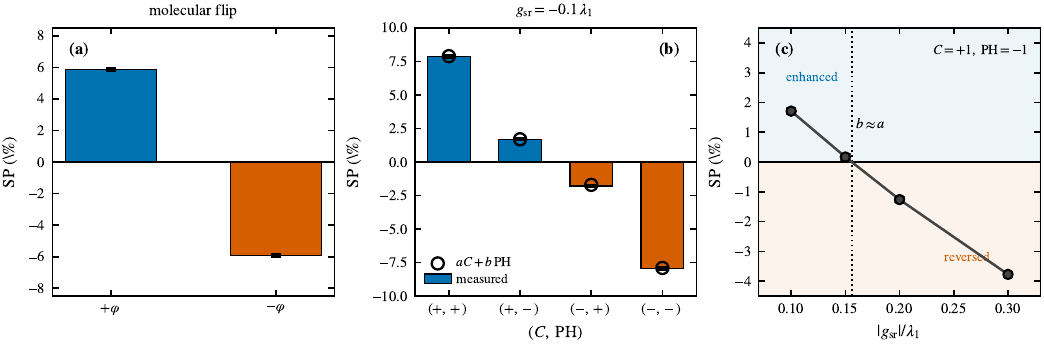}
  \caption{Effects of molecular chirality reversal, the four $(C,\PH)$
    combinations, and coupling strength on spin polarization.
    (a)~At $\gsr=0$, reversing $C$ gives polarizations of approximately equal
    magnitude and opposite sign in three independently sampled comparisons of
    the two enantiomers.
    (b)~Bars show the four $(C,\PH)$ combinations at
    $\gsr=-0.1\lambda_1$. Open circles show $\SP=aC+b\,\PH$, reconstructed
    from the same four values with $a=0.0483$ and $b=0.0308$; they are not
    independent predictions.
    (c)~For fixed $C=+1$ and $\PH=-1$, increasing $|\gsr|/\lambda_1$ reverses
    the sign of $\SP$.
    Error bars denote one standard deviation across independent ensemble
    estimates.
    \label{fig:main}}
\end{figure*}

\subsection{Effects of molecular and phonon chirality}
\label{sec:chirality-results}

We first compare the two molecular enantiomers at $\gsr=0$. Three independent
calculations consistently show that reversing $C$ leaves the magnitude of
$\SP$ approximately unchanged while reversing its sign
[Fig.~\ref{fig:main}(a)].

We next include $\Hsr$ and calculate $\SP$ for all four combinations of
$C,\PH\in\{+1,-1\}$. At $\gsr=-0.1\lambda_1$, the results are
\begin{equation}
\begin{aligned}
  \SP(+,+)&=+0.079, & \SP(+,-)&=+0.017,\\
  \SP(-,+)&=-0.018, & \SP(-,-)&=-0.079.
\end{aligned}
\label{eq:fourQuadrantValues}
\end{equation}
Simultaneous reversal of $C$ and $\PH$ approximately reverses the sign of
$\SP$ within the numerical resolution [Fig.~\ref{fig:main}(b)].

Applying the symmetry decomposition in Eq.~\eqref{eq:general4sign} to the
unrounded data gives
\begin{equation}
\begin{aligned}
  \alpha&=-0.0003, & a&=0.0483,\\
  b&=0.0308, & c&=0.0000.
\end{aligned}
\label{eq:extractedCoefficients}
\end{equation}
Both $\alpha$ and $c$ are close to zero, whereas $a$ and $b$ are clearly
nonzero, with $b/a=0.64$.

For fixed $C=+1$ and $\PH=-1$, we further vary $|\gsr|/\lambda_1$.
Increasing this ratio from 0.10 to 0.30 changes $\SP$ from $+0.017$ to
$-0.038$, with the sign reversal occurring between
$|\gsr|/\lambda_1=0.15$ and 0.20 [Fig.~\ref{fig:main}(c)].

\subsection{Intersite vibrational coupling and initial temperature distribution}
\label{sec:spatial-vibrations}

To examine the effect of intersite vibrational coupling, we set
$K_{\parallel}=5$~meV during propagation. Relative to the result at
$K_{\parallel}=0$, $|\SP|$ increases by a factor of about 1.7 for $c_f=0$
and by a factor of about 2 for $c_f=1$ (Fig.~S4). The initial vibrational
states in these calculations are still sampled from the site-factorized
distribution for $K_{\parallel}=0$.

A separate set of calculations keeps the mean initial sampling temperature at
$k_BT=25$~meV and changes only its spatial distribution along the molecule.
For $c_f=0$, placing the higher initial temperature on the injection side
increases $\SP$, whereas placing it on the absorbing side decreases $\SP$.
The two opposite initial-temperature profiles shift $\SP$ away from the
uniform-temperature result in opposite directions, and the average of the two
values is close to the result obtained from the uniform initial temperature
distribution (Fig.~S5).

No thermostat is used during propagation. The temperature profiles in
Fig.~S5 therefore determine only the initial nuclear coordinates and momenta.

As additional parameter checks, varying the CAP strength near
$\gamma_{\mathrm{CAP}}=25$~meV produces only a small change in $\SP$
(Fig.~S2). Varying $t_1$ from 4 to 16~meV changes $\SP$ by less than 10\%
(Fig.~S3).

\subsection{Steady-state NEGF results}
\label{sec:negf-comparison}

Supplementary Note~2 presents steady-state NEGF calculations with the same
form of spin-dependent hopping and a prescribed phonon angular momentum.
For the reference spin-independent contacts, the calculated spin polarization
is zero at zero phonon angular momentum. At nonzero angular momentum, the
four-sign decomposition gives a nonzero contribution that is odd under
reversal of its direction.

When the contacts preserve the mirror mapping, simultaneous reversal of
molecular chirality and the prescribed phonon angular momentum approximately
reverses the spin polarization. Introducing spin-dependent broadening at the
right contact breaks the mapping and produces nonzero constant and mixed
coefficients (Fig.~S7).

\section{Discussion}
\label{sec:discussion}

We first consider the case without $\Hsr$. Figures~\ref{fig:scans}(a) and
\ref{fig:scans}(b) show that the spin polarization is nearly zero when either
$\lambda_0=0$ or $\lambda_1=0$, indicating that both the static SOC at the
equilibrium geometry and its displacement-induced variation contribute to the
formation of spin polarization. At the same time, the dependence of $\SP$ on
$\lambda_1$ is nonmonotonic, so increasing the displacement-induced SOC does
not continuously enhance the spin polarization. As the temperature increases,
the initial distributions of nuclear coordinates and momenta broaden, and the
vibrational motion sampled by different trajectories becomes more diverse.
The smaller ensemble-averaged polarization may reflect increased cancellation
between spin-selective contributions, but the ensemble averages alone do not
establish this trajectory-level mechanism. The nonmonotonic dependence on
$\lambda_1$ and the decrease with temperature show that stronger vibrational
modulation does not necessarily enhance spin polarization. This temperature
trend is specific to the present preparation and dynamics: temperature can
affect electronic relaxation, interfacial spin response, and vibrational
populations differently in other CISS models~\cite{Alwan2023,Rudge2025}.

Ordinary molecular vibration can therefore affect spin polarization, but this
does not imply that phonon angular momentum directly acts on the electronic
spin. Figure~\ref{fig:scans}(d) shows that, at $\gsr=0$, changing the frequency
splitting between the two circularly polarized vibrational modes produces only
a small change in $\SP$. The splitting modifies the vibrational state and the
phonon angular momentum, but the phonon angular momentum does not directly
enter the electronic Hamiltonian in this limit. The electron responds only to
the displacement-induced changes in electronic hopping and SOC. Under the
investigated conditions, a nonzero phonon angular momentum without $\Hsr$
therefore produces only a small additional change in spin polarization.

When $\Hsr$ is included, phonon angular momentum directly modifies
spin-dependent electronic hopping, and the sign of this correction reverses
with the phonon angular momentum. The four $(C,\PH)$ combinations show that
reversing either molecular chirality or phonon chirality changes the final
spin polarization, while simultaneous reversal of $C$ and $\PH$ approximately
reverses the sign of $\SP$. The symmetry decomposition leaves the terms that
change sign with $C$ or $\PH$, whereas the constant and mixed $C\cdot\PH$
terms are nearly zero. The relation $\SP=aC+b\,\PH$ therefore describes the
symmetry dependence of the final result on the two chirality variables; it
does not represent two dynamically independent microscopic processes.

Figure~\ref{fig:main}(c) further shows that the molecular enantiomer alone does
not determine the sign of the spin polarization. With $C=+1$ and $\PH=-1$
fixed, varying only $|\gsr|/\lambda_1$ drives $\SP$ through zero and reverses
its sign. Increasing $|\gsr|$ strengthens the influence of phonon angular
momentum on spin-dependent hopping through $\Hsr$, thereby changing the
relative influence of molecular and phonon chirality on the final spin
polarization. When these effects favor opposite signs, changing the coupling
strength can change which effect dominates. Phonon angular momentum can
therefore tune the direction of spin polarization without changing the
molecular geometry. This control uses an imposed phonon bias rather than
predicting its generation from molecular handedness. Current-induced phonon
angular momentum~\cite{Nuomin2026} offers a possible physical source, but
its feedback on transport is not derived here. Likewise, the effective bond
coupling $\Hsr$ is distinct from the many-body and vibrationally mediated
exchange interactions in donor--bridge--acceptor models
~\cite{Chiesa2024,Chiesa2026}; those studies motivate related questions but
do not constitute a microscopic derivation of Eq.~\eqref{eq:HSR}.

Intersite coupling and the initial spatial distribution of molecular vibration
also affect spin polarization. The enhancement of $|\SP|$ at $K_{\parallel}>0$
persists even at $c_f=0$, so the enhancement does not require circular-mode
splitting. Because the initial distribution is sampled at $K_{\parallel}=0$,
switching on the positive intersite potential also changes the initial total
energy. This comparison therefore does not isolate the effect of spatial
correlations from the change in vibrational energy. The calculations with
spatially varying initial temperatures give a complementary result: at the same
mean temperature, placing the stronger initial vibration near the injection
side and near the absorbing side produces different spin polarizations. The
final spin polarization therefore depends on the initial spatial distribution
of molecular vibration, not only on the mean sampling temperature. Because no
thermal bath maintains the temperature profile during propagation, this
difference originates from the initial vibrational preparation rather than from
a steady-state temperature gradient.

The NEGF calculations provide an independent comparison using a different
transport description. In FSSH, the nuclear coordinates and momenta evolve in
time and the electron experiences a time-dependent electronic Hamiltonian; in
NEGF, the phonon angular momentum is fixed and the steady-state two-terminal
current is calculated. Time-dependent transport studies have also examined
spin-dependent propagation velocities and molecular occupancies
~\cite{Zhang2025,Stuermer2026}. Occupancy polarization is a different
observable from the collected-yield polarization used here. The NEGF benchmark shows that the prescribed
spin-dependent hopping produces a phonon-odd contribution in a steady-state
calculation as well as in the finite-time FSSH calculation. Reversing the
phonon chirality variable changes the sign of this contribution, but need not
reverse the total polarization at fixed molecular chirality. The agreement
concerns the chirality dependence and does not by itself validate the FSSH
approximation. The NEGF contact control further shows that the corresponding
mirror relation is a property of the full transport system: once the right
contact is modified so that the mirror relation is broken, the sign-reversal
relation under simultaneous reversal of molecular chirality and the prescribed
phonon angular momentum also disappears. The symmetry of the spin polarization
is therefore determined jointly by the molecule, the vibrational degree of
freedom, and the contacts. The qualitative agreement between FSSH and NEGF
supports the same role of the effective $\Hsr$ coupling in both transport
descriptions, whereas quantitative predictions for specific molecules or
materials will require more realistic electronic structures and phonon spectra,
together with spin--phonon coupling parameters derived from a microscopic
model.

\section{Conclusions}
\label{sec:conclusions}

We investigated how phonon angular momentum modifies CISS and separated its
contribution from that associated with molecular chirality. In the present
effective model, phonon angular momentum enters through spin-dependent
nearest-neighbor hopping and thereby provides a spin-selective control degree
of freedom distinct from molecular chirality.

The four $(C,\PH)$ combinations are consistent with
$\SP=aC+b\,\PH$. For preparation and collection protocols that preserve
the mirror mapping, the joint transformation
$(C,\PH,\SP)\to(-C,-\PH,-\SP)$ excludes both a chirality-independent
constant term and a mixed $C\cdot\PH$ term. The molecular- and
phonon-chirality components can therefore reinforce or cancel each other. In
particular, at fixed molecular chirality, increasing an opposing phonon
contribution can drive the total spin polarization through zero and reverse
its sign. Thus, within the present model, the CISS signal can be tuned and even
reversed through the phonon-rotation degree of freedom without changing the
molecular enantiomer.

\section*{Supplementary Material}
The supplementary material describes the numerical method, independent-ensemble
estimates, parameter controls, a comparison between FSSH and Ehrenfest dynamics,
and a two-terminal NEGF benchmark
(Figs.~S1--S7 and Table~S1).

\begin{acknowledgments}
W.D. acknowledges financial support from the National Natural Science
Foundation of China (Grant Nos.~22273075 and 22361142829) and the Zhejiang
Provincial Natural Science Foundation (Grant No.~XHD24B0301). The authors
acknowledge computational resources and technical support from the
High-Performance Computing Center at Westlake University. Amikam Levy was
supported by the ISF under Grant No.~3105/23.
\end{acknowledgments}

\section*{Author Declarations}
\subsection*{Conflict of Interest}
The authors have no conflicts to disclose.

\subsection*{Author Contributions}
\textbf{Shi-Qi Zhang}: Conceptualization (equal); Formal analysis (lead);
Investigation (lead); Methodology (equal); Software (lead); Visualization
(lead); Writing -- original draft (lead). \textbf{Vipul Upadhyay}: Formal
analysis (supporting); Methodology (supporting); Software (supporting);
Visualization (supporting); Writing -- review \& editing (supporting).
\textbf{Jiayue Han}: Formal analysis (supporting); Methodology (supporting);
Software (supporting); Validation (lead); Writing -- review \& editing
(supporting). \textbf{Amikam Levy}: Conceptualization (supporting); Methodology
(supporting); Supervision (supporting); Writing -- review \& editing
(supporting). \textbf{Wenjie Dou}: Conceptualization (equal); Funding
acquisition (lead); Project administration (lead); Supervision (lead); Writing
-- review \& editing (lead).

\section*{Data Availability}
The data that support the reported findings are available from the
corresponding author upon reasonable request.

\appendix

\section{Circular-mode basis and phonon angular momentum}
\label{app:modemixing}

To relate the circular-mode splitting to the initial phonon angular momentum,
we diagonalize the local vibrational Hamiltonian for $K_{\parallel}=0$.
Define
\begin{equation}
  b_{j,\alpha}=\frac{R_{j,\alpha}+iP_{j,\alpha}}{\sqrt2},
  \qquad
  \Phi_j=(b_{j,1},b_{j,2})^{\!\top}.
  \label{eq:localBosons}
\end{equation}
Using $\bm\tau=(\tau^x,\tau^y,\tau^z)$ for the Pauli matrices in the
two-mode space and $\tau^0$ for the $2\times2$ identity matrix, the local
vibrational Hamiltonian becomes
\begin{equation}
  H_{0,j}+H_{\Omega,j}
  =\hbar\Phi_j^\dagger(\omega_0\tau^0+\Omega\tau^y)\Phi_j
  +\mathrm{const}.
  \label{eq:phononMatrix}
\end{equation}
The matrix $\tau^y$ also determines the phonon angular momentum:
\begin{equation}
\begin{aligned}
  \Phi_j^\dagger\tau^y\Phi_j
  &=-i\left(b_{j,1}^\dagger b_{j,2}
  -b_{j,2}^\dagger b_{j,1}\right)\\
  &=R_{j,1}P_{j,2}-R_{j,2}P_{j,1}\\
  &=L_{\mathrm{ph},j}.
\end{aligned}
\label{eq:tauyL}
\end{equation}

The same circular basis therefore resolves both the energy and angular
momentum. The operators
\begin{equation}
  b_{j,\pm}=\frac{b_{j,1}\mp ib_{j,2}}{\sqrt2}
  \label{eq:circularOperators}
\end{equation}
diagonalize Eq.~\eqref{eq:phononMatrix}. In the circular basis,
\begin{equation}
  L_{\mathrm{ph},j}
  =b_{j,+}^\dagger b_{j,+}-b_{j,-}^\dagger b_{j,-}.
  \label{eq:Lnumber}
\end{equation}
In this basis, the local Hamiltonian is
\begin{equation}
\begin{aligned}
  H_{0,j}+H_{\Omega,j}={}&
  \hbar(\omega_0+\Omega)b_{j,+}^\dagger b_{j,+}\\
  &+\hbar(\omega_0-\Omega)b_{j,-}^\dagger b_{j,-}
  +\mathrm{const}.
\end{aligned}
\label{eq:Hdiag}
\end{equation}
The two circular frequencies are $\omega_\pm=\omega_0\pm\Omega$. A quantum
in the $+$ mode carries angular momentum $+\hbar$, and a quantum in the $-$
mode carries angular momentum $-\hbar$, as follows from
Eq.~\eqref{eq:Lnumber}.

Because the two modes carry opposite angular momenta, their population
imbalance determines the thermal mean. At temperature $T>0$, the occupations
are
$\avg{b_{j,\pm}^\dagger b_{j,\pm}}=n_B(\omega_\pm)$, where
$n_B(\omega)=[\exp(\hbar\omega/k_BT)-1]^{-1}$. Hence
\begin{equation}
\begin{aligned}
  \avg{\Lph}&=n_B(\omega_+)-n_B(\omega_-)\\
  &\xrightarrow{\,k_BT\gg\hbar\omega_0\,}
  -\frac{2\Omega k_BT}{\hbar(\omega_0^2-\Omega^2)}\\
  &\xrightarrow{\,|\Omega|\ll\omega_0\,}
  -\frac{2\Omega k_BT}{\hbar\omega_0^2}.
\end{aligned}
\label{eq:LhighT}
\end{equation}
To connect this result to the phonon initialization used in the dynamics,
we write the thermal Wigner variances as
\begin{equation}
  w_\pm^2=\frac12\coth\!\left(\frac{\hbar\omega_\pm}{2k_BT}\right)
  =n_B(\omega_\pm)+\frac12.
  \label{eq:WignerBose}
\end{equation}
The zero-point contributions cancel in $w_+^2-w_-^2$. Thermal Wigner sampling
therefore gives the same mean angular momentum as the Bose occupations.
For $T>0$, $n_B(\omega)$ decreases with frequency, so $\Omega>0$ gives
$\avg{\Lph}<0$. Thus, for the geometry and mode convention used here,
\begin{equation}
  \sgn\avg{\Lph}=-\sgn\Omega=-\PH,
  \qquad T>0,\quad \Omega\ne0.
  \label{eq:exactSign}
\end{equation}
At $T=0$, both Bose occupations vanish, and the mean angular momentum is zero.
These expressions describe the uncoupled-site thermal distribution and are
not identities for the nonequilibrium phonon distribution during electronic
transport.

\bibliographystyle{aipnum4-2}
\bibliography{references}

\end{document}


\title{Supplementary Material for\\
Phonon chirality as an additive control of CISS: a symmetry-protected law}

\author{Shi-Qi Zhang}
\affiliation{Key Laboratory for Quantum Materials of Zhejiang Province,
Department of Physics, School of Science and Research Center for Industries of
the Future, Westlake University, Hangzhou, Zhejiang 310030, China}

\author{Vipul Upadhyay}
\affiliation{Department of Chemistry, Bar-Ilan University, Ramat-Gan 52900, Israel}
\affiliation{Institute of Nanotechnology and Advanced Materials, Bar-Ilan University,
Ramat-Gan 52900, Israel}
\affiliation{Center for Quantum Entanglement Science and Technology, Bar-Ilan University,
Ramat-Gan 52900, Israel}

\author{Jiayue Han}
\affiliation{Department of Chemistry, School of Science and Research Center for
Industries of the Future, Westlake University, Hangzhou, Zhejiang 310030, China}
\affiliation{Institute of Natural Sciences, Westlake Institute for Advanced
Study, Hangzhou, Zhejiang 310024, China}

\author{Amikam Levy}
\affiliation{Department of Chemistry, Bar-Ilan University, Ramat-Gan 52900, Israel}
\affiliation{Institute of Nanotechnology and Advanced Materials, Bar-Ilan University,
Ramat-Gan 52900, Israel}
\affiliation{Center for Quantum Entanglement Science and Technology, Bar-Ilan University,
Ramat-Gan 52900, Israel}

\author{Wenjie Dou}
\email{douwenjie@westlake.edu.cn}
\affiliation{Department of Chemistry, School of Science and Research Center for
Industries of the Future, Westlake University, Hangzhou, Zhejiang 310030, China}
\affiliation{Institute of Natural Sciences, Westlake Institute for Advanced
Study, Hangzhou, Zhejiang 310024, China}
\affiliation{Key Laboratory for Quantum Materials of Zhejiang Province,
Department of Physics, School of Science and Research Center for Industries of
the Future, Westlake University, Hangzhou, Zhejiang 310030, China}

\date{\today}

\maketitle

\section*{Supplementary Note 1: Numerical implementation of the FSSH dynamics}

\subsection*{Overall propagation scheme}

The adiabatic surfaces used in the FSSH dynamics are defined by
\begin{equation}
  H_{\mathrm{SH}}(\bm R)
  =H_{\mathrm{el}}+H_{\mathrm{el-ph}}(\bm R).
  \label{eq:SHSH}
\end{equation}
For each nuclear configuration \(\bm R\), \(H_{\mathrm{SH}}\) is diagonalized,
and one adiabatic state or exactly degenerate subspace is designated as
active. States within an exactly degenerate subspace are not separate hopping
targets. The electronic density matrix retains all $2N$ states.

When the effective spin--phonon coupling is included, the electronic density
matrix is propagated with
\begin{equation}
  H_{\mathrm e}=H_{\mathrm{SH}}+\Hsr.
  \label{eq:SHeFull}
\end{equation}
The interaction \(\Hsr\) does not redefine the adiabatic surfaces. Its
mean-field contribution to the classical motion is evaluated with the
normalized surviving electronic density matrix
\begin{equation}
  \widetilde\rho=\frac{\rho}{\operatorname{Tr}\rho},
  \qquad \operatorname{Tr}\rho>0,
  \label{eq:Snormalizedrho}
\end{equation}
and its off-diagonal contribution in the \(H_{\mathrm{SH}}\) adiabatic basis is
included directly in the FSSH hopping rate.

A complex absorbing potential (CAP) at the terminal site removes electronic
population that reaches the absorbing end. The CAP is used as an absorbing
sink and to evaluate the transmitted spin-resolved yields; it is not included
as an additional contribution to the hopping rate.

\subsection*{Nuclear equations of motion}

Atomic units ($\hbar=1$) are used in Supplementary Note~1.
Each site contains two transverse vibrational coordinates
\((R_{j,1},R_{j,2})\) and their conjugate momenta \((P_{j,1},P_{j,2})\). We
write the classical equations as
\begin{equation}
\begin{aligned}
  \dot{\bm R}&=\dot{\bm R}_{\mathrm{bare}}
  +\frac{\partial\langle\Hsr\rangle_{\widetilde\rho}}{\partial\bm P},\\
  \dot{\bm P}&=\dot{\bm P}_{\mathrm{bare}}
  -\frac{\partial\langle\Hsr\rangle_{\widetilde\rho}}{\partial\bm R},
\end{aligned}
\label{eq:SphononEOM}
\end{equation}
where
\begin{equation}
  \langle\Hsr\rangle_{\widetilde\rho}
  =\operatorname{Tr}(\widetilde\rho\Hsr).
  \label{eq:SHSRmeanfield}
\end{equation}
The bare equations, before the \(\Hsr\) mean-field contribution is added, are
\begin{equation}
\begin{aligned}
  \dot R^{\mathrm{bare}}_{j,1}&=\omega_0P_{j,1}-\Omega R_{j,2},\\
  \dot R^{\mathrm{bare}}_{j,2}&=\omega_0P_{j,2}+\Omega R_{j,1},\\
  \dot P^{\mathrm{bare}}_{j,1}&=-\omega_0R_{j,1}-\Omega P_{j,2}
  +F^{\mathrm{SH}}_{j,1}+F^{\parallel}_{j,1},\\
  \dot P^{\mathrm{bare}}_{j,2}&=-\omega_0R_{j,2}+\Omega P_{j,1}
  +F^{\mathrm{SH}}_{j,2}+F^{\parallel}_{j,2}.
\end{aligned}
\label{eq:SphononEOMbare}
\end{equation}
Here \(F^{\mathrm{SH}}_{j,\alpha}\) is the force on the current active
adiabatic state or degenerate subspace. Because \(\Hsr\) depends on \(\bm P\), the first line of
Eq.~\eqref{eq:SphononEOM} changes the actual nuclear velocity used in the FSSH
hopping rate. When the intersite vibrational coupling \(K_{\parallel}\) is
included, the additional force is
\begin{equation}
F^{\parallel}_{j,\alpha}=
\begin{cases}
 K_{\parallel}(R_{2,\alpha}-R_{1,\alpha}), & j=1,\\[2pt]
 K_{\parallel}(R_{j-1,\alpha}+R_{j+1,\alpha}-2R_{j,\alpha}),
 & 1<j<N,\\[2pt]
 K_{\parallel}(R_{N-1,\alpha}-R_{N,\alpha}), & j=N.
\end{cases}
\label{eq:SFparallel}
\end{equation}
The baseline calculations use \(K_{\parallel}=0\); the force in
Eq.~\eqref{eq:SFparallel} is included only in the corresponding
\(K_{\parallel}\) calculations.

No explicit Berry-curvature (geometric magnetic) force is added to these
nuclear equations. Such corrections have been developed for surface hopping
with complex Hamiltonians~\cite{MiaoSubotnik2019}. Complex spin--orbit coupling
is retained in the electronic Hamiltonian and nonadiabatic couplings, but
this does not include the associated geometric force on the nuclei. The
mean-field derivatives of $\Hsr$ in Eq.~\eqref{eq:SphononEOM} are retained;
they are not an added Berry-force correction. Omitting the latter is an
approximation, not an assumption that the Berry curvature vanishes.

\subsection*{Surface hopping and momentum adjustment}

For each nuclear configuration,
\begin{equation}
  H_{\mathrm{SH}}U=U\,\operatorname{diag}(E_1,\ldots,E_{2N})
\end{equation}
is diagonalized. For an individual eigenstate $a$, the surface force is
\begin{equation}
  F^{\mathrm{SH}}_{j,\alpha}(a)
  =-\left\langle\phi_a\left|
  \frac{\partial H_{\mathrm{SH}}}{\partial R_{j,\alpha}}
  \right|\phi_a\right\rangle.
  \label{eq:Ssurfaceforce}
\end{equation}
An exactly degenerate subspace $A$ containing $m_A$ states has energy
$E_A=m_A^{-1}\sum_{a\in A}E_a$ and force
$F^{\mathrm{SH}}_{j,\alpha}(A)=m_A^{-1}\sum_{a\in A}
F^{\mathrm{SH}}_{j,\alpha}(a)$. A nondegenerate state is the special case
$m_A=1$. The equal-weight subspace force avoids selecting an arbitrary
basis vector within a degeneracy.
The nonadiabatic coupling vector between two different adiabatic states is
denoted by
\begin{equation}
  \bm d_{ab}=\langle\phi_a|\nabla_{\bm R}\phi_b\rangle.
  \label{eq:Snacdef}
\end{equation}

In the \(H_{\mathrm{SH}}\) adiabatic basis,
\begin{equation}
  \widetilde\rho^{\mathrm{ad}}=U^\dagger\widetilde\rho U,
  \qquad
  H_{\mathrm{SR}}^{\mathrm{ad}}=U^\dagger\Hsr U.
  \label{eq:Sadiabaticquantities}
\end{equation}
For states $a$ and $b$ belonging to different adiabatic subspaces, the
pairwise population-transfer numerator contains the usual nonadiabatic term evaluated with the actual
nuclear velocity and the direct population-transfer contribution from
\(\Hsr\):
\begin{equation}
\begin{aligned}
  \Phi_{a\rightarrow b}={}&
  2\,\operatorname{Re}\!\left[
    \widetilde\rho^{\mathrm{ad}}_{ba}
    \dot{\bm R}\!\cdot\!\bm d_{ab}\right]
  +2\,\operatorname{Im}\!\left[
    \widetilde\rho^{\mathrm{ad}}_{ab}
    (H_{\mathrm{SR}}^{\mathrm{ad}})_{ba}\right].
\end{aligned}
\label{eq:Spairflux}
\end{equation}
Because the velocity in Eq.~\eqref{eq:Spairflux} is the full velocity in
Eq.~\eqref{eq:SphononEOM}, its nonadiabatic contribution can equivalently be
written as a bare part plus the indirect velocity correction generated by
\(\Hsr\),
\begin{equation}
  \Phi^{\mathrm{NAC}}_{a\rightarrow b}
  =\Phi^{\mathrm{NAC,bare}}_{a\rightarrow b}
  +\Phi^{\mathrm{HSR,velocity}}_{a\rightarrow b}.
  \label{eq:SHSRvelocityflux}
\end{equation}
For an active subspace $A$ and target subspace $B$, the signed pairwise
contributions are summed before taking the positive part. With
$p_A=\sum_{a\in A}\widetilde\rho^{\mathrm{ad}}_{aa}$, the rate and one-step
hopping probability are
\begin{equation}
  k_{A\rightarrow B}
  =\max\!\left[0,
  \frac{\sum_{a\in A,b\in B}\Phi_{a\rightarrow b}}{p_A}
  \right],
  \qquad
  g_{A\rightarrow B}=k_{A\rightarrow B}\Delta t.
  \label{eq:Sstateprob}
\end{equation}
For nondegenerate states, this expression reduces to the usual state-to-state
rate.
For \(\gsr=0\), Eq.~\eqref{eq:Spairflux} reduces to the standard FSSH
population-transfer expression.

The momentum-rescaling direction must be real even though the derivative
couplings are complex. For a proposed $A\rightarrow B$ hop, the implementation
first contracts the couplings with the electronic density matrix,
\begin{equation}
  \bm q_{AB}=\sum_{a\in A,b\in B}
  \widetilde\rho^{\mathrm{ad}}_{ba}\bm d_{ab}.
  \label{eq:Scomplexdirection}
\end{equation}
This contraction is invariant under independent unitary basis rotations
within the two subspaces. We then select a phase that maximizes the squared
norm of the real part, following the phase-maximization prescription for
complex couplings~\cite{MiaoSubotnik2019}, here applied to $\bm q_{AB}$:
\begin{equation}
  \theta_*\in\operatorname*{arg\,max}_{\theta}
  \left\|\operatorname{Re}(e^{-i\theta}\bm q_{AB})\right\|^2,
  \qquad
  \bm n_{AB}=\operatorname{Re}(e^{-i\theta_*}\bm q_{AB}).
  \label{eq:Srealdirection}
\end{equation}
Writing $\bm q_{AB}=\bm x+i\bm y$, the phase is obtained from
\begin{equation}
  \theta_0=\frac12\operatorname{atan2}
  \!\left(2\bm x\!\cdot\!\bm y,\|\bm x\|^2-\|\bm y\|^2\right).
  \label{eq:Sdirectionphase}
\end{equation}
The code compares the real vectors associated with $\theta_0$ and
$\theta_0+\pi/2$ and retains the one with the larger norm. Thus the procedure
is not simply $\operatorname{Re}\bm d_{ab}$ in an arbitrary electronic
phase convention. For a nonzero direction, we define
$\hat{\bm d}_{AB}=\bm n_{AB}/\|\bm n_{AB}\|$ and adjust the momentum as
\begin{equation}
  \bm P'=\bm P+\kappa\hat{\bm d}_{AB},
  \label{eq:Smomentumupdate}
\end{equation}
where \(\kappa\) is chosen to conserve the energy including the vibrational
Hamiltonian, the active \(H_{\mathrm{SH}}\) surface, and the conditional
mean-field contribution of \(\Hsr\),
\begin{equation}
\begin{aligned}
  &H_{\mathrm{vib}}(\bm R,\bm P')+E_B
  +\langle\Hsr(\bm R,\bm P')\rangle_{\widetilde\rho}\\
  ={}&H_{\mathrm{vib}}(\bm R,\bm P)+E_A
  +\langle\Hsr(\bm R,\bm P)\rangle_{\widetilde\rho}.
\end{aligned}
  \label{eq:Shopenergy}
\end{equation}
The electronic density matrix and nuclear coordinates are held fixed during
this adjustment. Of the two real roots, the root with smaller $|\kappa|$ is
used. If the coupling supplies no nonzero real direction, a hop with a
nonzero subspace energy gap is rejected rather than assigned an unrelated
rescaling direction.
If Eq.~\eqref{eq:Shopenergy} has no real solution, the hop is frustrated. The
production calculations use the \emph{reject} prescription: the active state
and the nuclear momentum are both left unchanged.

\subsection*{Electronic propagation, CAP absorption, and spin polarization}

The CAP part of the electronic propagation is separated from the remaining
electronic evolution using a second-order Strang splitting,
\begin{equation}
\begin{aligned}
  \rho(t+\Delta t)&\simeq
  e^{\mathcal L_D\Delta t/2}
  e^{\mathcal L_U\Delta t}
  e^{\mathcal L_D\Delta t/2}\rho(t),\\
  \mathcal L_U\rho&=-i[H_{\mathrm e},\rho],\\
  \mathcal L_D\rho&=-\frac12\{\Gcap,\rho\}.
\end{aligned}
\label{eq:Sstrang}
\end{equation}
Because \(\Gcap\) is diagonal in the site--spin basis, each CAP half-step can
be propagated analytically. Between the two CAP half-steps, \(\bm R\),
\(\bm P\), and \(\rho\) are propagated numerically with the same time step as the FSSH
propagation.

The spin-resolved CAP current at the absorbing end is
\begin{equation}
  \mathcal J_s(t)=\gamma_{\mathrm{CAP}}\rho_{Ns,Ns}(t),
\end{equation}
and the accumulated transmitted yield is
\begin{equation}
  J_s=\int_0^{t_f}\mathcal J_s(t)\,\mathrm dt.
  \label{eq:SCAPyield}
\end{equation}
The propagation obeys the population balance
\begin{equation}
  \operatorname{Tr}\rho(t_f)+J_\uparrow+J_\downarrow\simeq1.
  \label{eq:SCAPbalance}
\end{equation}

\section*{Supplementary figures and method controls}

\begin{table}[htbp]
  \caption{Point estimates from the exact four-point parity decomposition
    $\SP=\alpha+aC+b\,\PH+cC\cdot\PH$ at $\gsr=-0.1\lambda_1$, obtained using
    FSSH/mean-field and Ehrenfest nuclear dynamics.}
  \label{tab:methodcompare}
  \begin{ruledtabular}
  \begin{tabular}{lccccc}
    dynamical method & $a$ & $b$ & $c$ & $\alpha$ & $b/a$ \\
    \colrule
    FSSH/mean-field & $0.0483$ & $0.0308$ & $0.0000$ & $-0.0003$ & $0.64$ \\
    Ehrenfest dynamics
      & $0.0429$ & $0.0277$ & $0.0011$ & $0.0003$ & $0.65$ \\
  \end{tabular}
  \end{ruledtabular}
\end{table}

\begin{figure}[htbp]
  \centering
  \includegraphics[width=8.6cm]{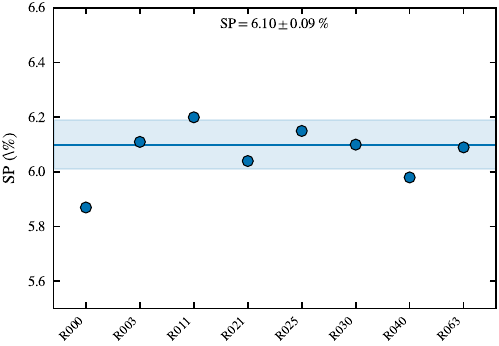}
  \caption{Variation across independent ensemble estimates at $\gsr=0$ and
    fixed physical parameters. The mean and standard deviation are
    $\SP=0.0610\pm0.0009$.}
  \label{fig:S1}
\end{figure}

\begin{figure}[htbp]
  \centering
  \includegraphics[width=8.6cm]{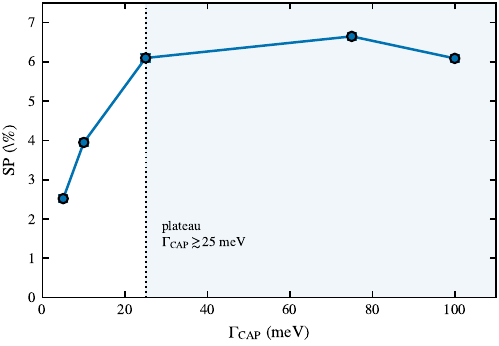}
  \caption{Transmitted spin polarization as a function of the CAP strength
    $\gamma_{\mathrm{CAP}}$. The polarization approaches a plateau over a
    range that includes the baseline choice $\gamma_{\mathrm{CAP}}=25$~meV.}
  \label{fig:S2}
\end{figure}

\begin{figure}[htbp]
  \centering
  \includegraphics[width=8.6cm]{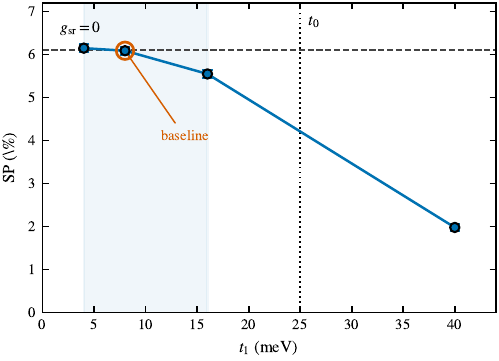}
  \caption{Transmitted spin polarization as a function of the
    spin-independent hopping-modulation strength $t_1$ at $\gsr=0$. The
    polarization changes by less than 10\% for $t_1=4$--$16$~meV and decreases
    more strongly when $t_1$ approaches $t_0=25$~meV.}
  \label{fig:S3}
\end{figure}

\begin{figure}[htbp]
  \centering
  \includegraphics[width=8.6cm]{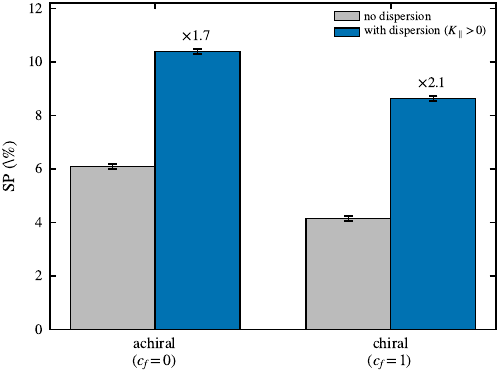}
  \caption{Coupling-quench control for the intersite force constant
    $K_{\parallel}$. Relative to $K_{\parallel}=0$, setting
    $K_{\parallel}=5$~meV during propagation increases $|\SP|$ by factors of
    approximately 1.7 and 2 for $c_f=0$ and $c_f=1$, respectively. The initial
    distribution remains the site-factorized $K_{\parallel}=0$ Wigner
    distribution. Because an enhancement is also present at $c_f=0$, it is not
    uniquely attributable to the circular-mode bias.}
  \label{fig:S4}
\end{figure}

\begin{figure}[htbp]
  \centering
  \includegraphics[width=8.6cm]{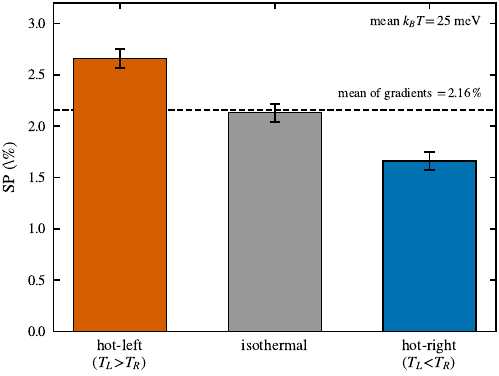}
  \caption{Site-dependent initial-temperature-profile control at mean
    sampling temperature $k_BT=25$~meV and $c_f=0$. Assigning the higher
    initial temperature to the injection side and to the absorbing side shifts
    $\SP$ in opposite directions. The arithmetic mean of the two profile
    orientations is close to the isothermal value. No thermostat maintains the
    profile during propagation.}
  \label{fig:S5}
\end{figure}

\begin{figure}[htbp]
  \centering
  \includegraphics[width=8.6cm]{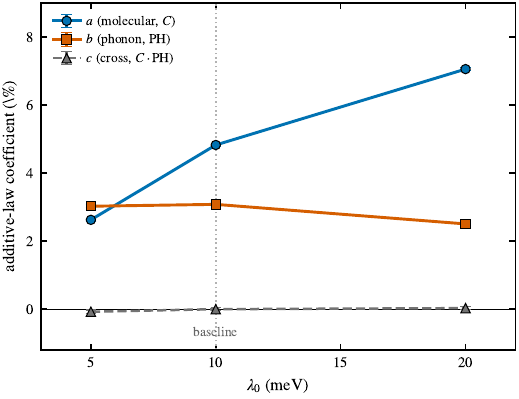}
  \caption{Parity coefficients obtained from the four $(C,\PH)$ combinations as
    functions of $\lambda_0$ at $\gsr=-0.1\lambda_1$ and $c_f=1$.
    Increasing $\lambda_0$ from 5 to 20~meV increases $a$ by a factor of 2.7.
    Over the same interval, $b$ changes by less than 20\%, and the extracted
    values of $c$ remain close to zero.}
  \label{fig:S6}
\end{figure}

\clearpage
\section*{Supplementary Note 2: Steady-state NEGF benchmark with a prescribed phonon angular momentum}

The FSSH calculation follows dynamical phonons and collects electronic
population over a finite propagation time. To examine the same assumed
spin-dependent hopping in a steady-state setting, the NEGF benchmark instead
imposes a uniform mean phonon angular momentum and connects the molecule to
left and right leads. The reference configuration has spin-independent
lead couplings; the right-contact control in Fig.~\ref{fig:S7}(d) introduces
spin dependence. The phonon coordinates are not
propagated, and the observable is a two-terminal current rather than
time-integrated CAP absorption.

To compare the two calculations, the angular-momentum conventions must first
be related. Here $\eta=\pm1$ specifies the direction of the prescribed mean
phonon angular momentum when its magnitude is nonzero. In the baseline FSSH
thermal preparation at $T>0$ and $\Omega\ne0$, $\PH=\sgn\Omega$ and
$\sgn\avg{\Lph}=-\PH$. The same initial angular-momentum direction therefore
corresponds to $\eta=-\PH$.

\subsection*{Mean-field Hamiltonian}

With the angular momentum prescribed, the molecular Hamiltonian contains the
static electronic term and its mean-field spin--phonon coupling:
\begin{equation}
  H^{\mathrm{MF}}_{\mathrm{mol}}(C,\eta)
  =H_{\mathrm{el}}(C)+H^{\mathrm{MF}}_{\mathrm{SR}}(\eta).
  \label{eq:SnegfH}
\end{equation}
The static term $H_{\mathrm{el}}(C)$ is the same as in the main text. The
mean-field term is obtained by imposing the uniform bond angular momentum
\begin{equation}
  \avg{\bar L_{\mathrm{ph},j}}=L_0\eta,
  \qquad L_0\ge0.
  \label{eq:SnegfL0}
\end{equation}
Substitution into the spin--phonon coupling gives
\begin{equation}
  H^{\mathrm{MF}}_{\mathrm{SR}}(\eta)
  =-\gsr\sum_{j=1}^{N-1}\sum_{s=\pm1}
  sL_0\eta\,\hat T_{j,s},
  \label{eq:SnegfHSR}
\end{equation}
where
$\hat T_{j,s}=c^\dagger_{js}c_{j+1,s}+c^\dagger_{j+1,s}c_{js}$.
The resulting spin-dependent nearest-neighbor hopping amplitude is
\begin{equation}
  t_s(\eta)=t_0+s\gsr L_0\eta.
  \label{eq:Snegfhop}
\end{equation}
Reversing $\eta$ exchanges the hopping corrections for the two spin states.
At $L_0=0$, $\eta$ has no effect on the Hamiltonian. For fixed $L_0>0$ and
$\eta$, the mean-field coupling is odd under electronic time reversal.
Time reversal maps the imposed angular momentum $L_0\eta$ to $-L_0\eta$;
the invariance of the full spin--phonon interaction does not make the
electronic Hamiltonian invariant at fixed nonzero angular momentum.

\subsection*{Green function and spin-resolved current}

To compute the current through this mean-field molecule, we include the two
leads through their self-energies. In the wide-band approximation, the
lead-coupling matrices are energy independent, and the retarded Green function
is
\begin{equation}
  G(E;C,\eta)=\left[E I-H^{\mathrm{MF}}_{\mathrm{mol}}(C,\eta)
  -\Sigma_L(E)-\Sigma_R(E)\right]^{-1}.
  \label{eq:SnegfG}
\end{equation}
The lead self-energies are
\begin{equation}
  \Sigma_L=-\frac{i}{2}\Gamma_L,
  \qquad
  \Sigma_R=-\frac{i}{2}\Gamma_R,
  \label{eq:SnegfSigma}
\end{equation}
where $\Gamma_L$ and $\Gamma_R$ are the lead-coupling matrices in the molecular
site--spin basis. Both matrices are spin independent in the reference
configuration, whereas $\Gamma_R$ is spin dependent in the contact control
specified below. The projector $P_{R,s}$ selects spin $s$
at the right contact. The spin-resolved transmission function is
\begin{equation}
  T_s(E;C,\eta)=\operatorname{Tr}\!\left[
  P_{R,s}\Gamma_R G(E;C,\eta)\Gamma_LG^\dagger(E;C,\eta)
  \right].
  \label{eq:SnegfT}
\end{equation}
The spin-resolved current entering the right lead is
\begin{equation}
  I_{R,s}(C,\eta)=\frac{e}{h}\int dE\,
  T_s(E;C,\eta)\left[f_L(E)-f_R(E)\right].
  \label{eq:SnegfI}
\end{equation}
Here $f_L$ and $f_R$ are the lead Fermi distributions, $e$ is the elementary
charge, and $h$ is Planck's constant. The two-terminal spin polarization is
\begin{equation}
  \SP_{2T}(C,\eta)=
  \frac{I_{R,\uparrow}(C,\eta)-I_{R,\downarrow}(C,\eta)}
       {I_{R,\uparrow}(C,\eta)+I_{R,\downarrow}(C,\eta)}.
  \label{eq:SnegfSP}
\end{equation}

\begin{figure}[htbp]
  \centering
  \includegraphics[width=0.95\textwidth]{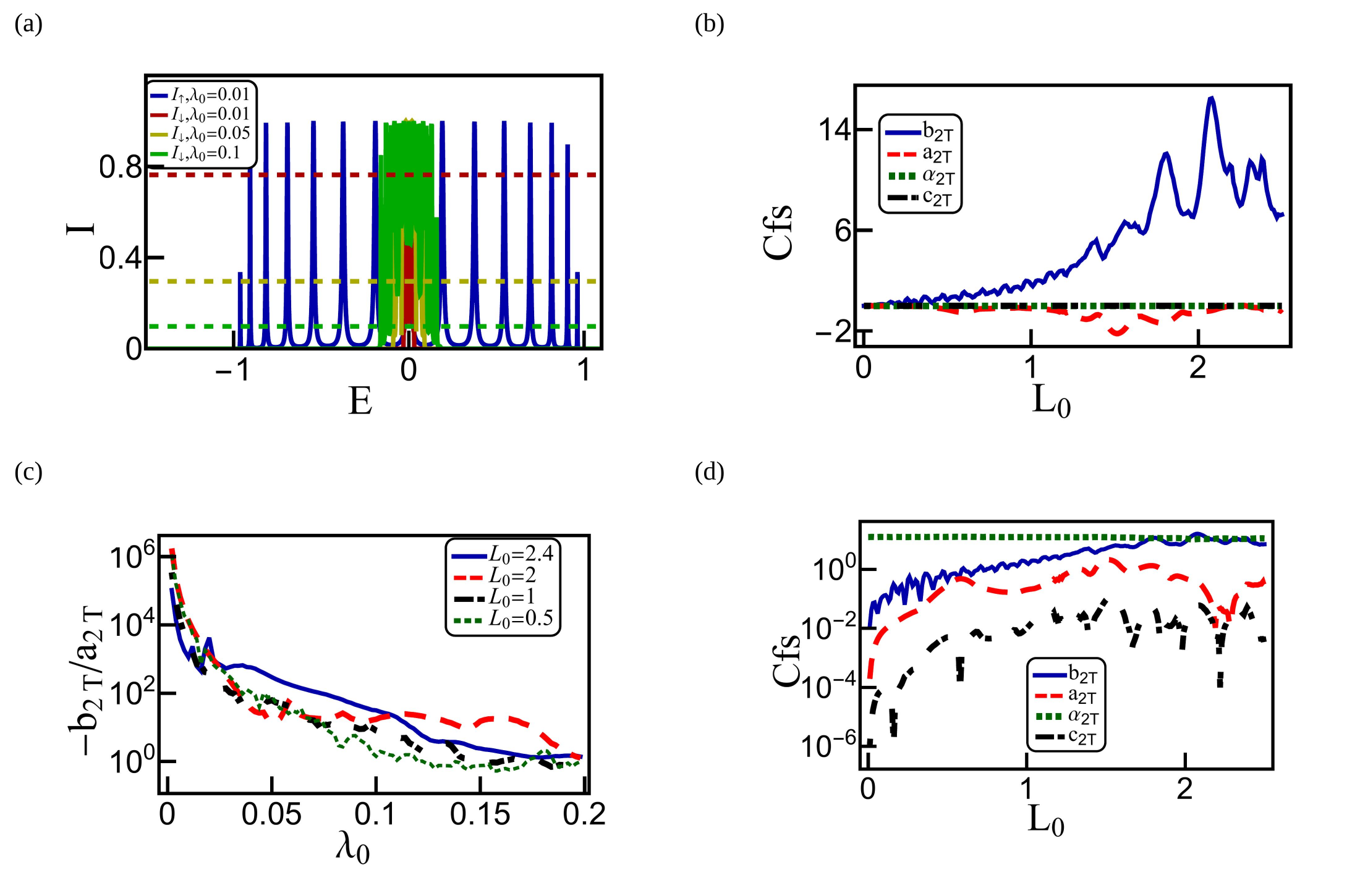}
  \caption{Steady-state NEGF benchmark with a prescribed mean phonon angular
    momentum. (a) Spin-resolved transmission spectra for $C=\eta=+1$.
    (b) Exact four-sign coefficients as functions of $L_0$.
    (c) The ratio $-b_{2T}/a_{2T}$ as a function of $\lambda_0$.
    (d) The corresponding coefficients after the spin-asymmetric right-contact
    modification in Eq.~\eqref{eq:SnegfBrokenContact}, which breaks the mirror
    mapping. The coefficients in panels (b) and (d) are expressed in percent,
    i.e., the plotted values are 100 times the dimensionless coefficients in
    Eq.~\eqref{eq:Snegfdecomp}. ``Cfs'' denotes coefficients.}
  \label{fig:S7}
\end{figure}

For Fig.~\ref{fig:S7}, energies are expressed in the unit $E_0=0.1$~eV.
Unless a panel varies the corresponding parameter, the NEGF calculation uses
$t_0=0.25E_0$, $\varepsilon_0=0$, $\lambda_0=0.10E_0$,
$\Gamma_0=t_0/4$, and $\gsr=0.10E_0$. The model helix has three turns and
five sites per turn. Panel~(d) breaks the reference mirror mapping through the
spin-asymmetric right-contact broadenings
\begin{equation}
  \Gamma_{R,\uparrow}=\Gamma_0(1+p),\qquad
  \Gamma_{R,\downarrow}=\Gamma_0(1-p),\qquad p=0.25.
  \label{eq:SnegfBrokenContact}
\end{equation}

\subsection*{Decomposition over the four \texorpdfstring{$(C,\eta)$}{(C, eta)} combinations}

To compare the chirality dependence with the FSSH results, we apply the same
four-sign decomposition to the NEGF polarizations. For
$C,\eta\in\{+1,-1\}$,
\begin{equation}
  \SP_{2T}(C,\eta)=
  \alpha_{2T}+a_{2T}C+b_{2T}\eta+c_{2T}\,C\cdot\eta.
  \label{eq:Snegfdecomp}
\end{equation}
Let $\SP_{++}$, $\SP_{+-}$, $\SP_{-+}$, and $\SP_{--}$ denote the
polarizations for the four sign combinations of $(C,\eta)$. The coefficients
are
\begin{align}
  \alpha_{2T}&=\tfrac14
  (\SP_{++}+\SP_{+-}+\SP_{-+}+\SP_{--}),\\
  a_{2T}&=\tfrac14
  (\SP_{++}+\SP_{+-}-\SP_{-+}-\SP_{--}),\\
  b_{2T}&=\tfrac14
  (\SP_{++}-\SP_{+-}+\SP_{-+}-\SP_{--}),\\
  c_{2T}&=\tfrac14
  (\SP_{++}-\SP_{+-}-\SP_{-+}+\SP_{--}).
  \label{eq:Snegfcoeffs}
\end{align}
This decomposition separates the term $b_{2T}\eta$, which is even in $C$
and odd in $\eta$, from the mixed term $c_{2T}\,C\cdot\eta$, which is odd in
both labels. The contact geometry determines whether the same mirror
constraint applies as in the FSSH model. When the contacts and Hamiltonian
obey the joint mirror mapping,
\begin{equation}
  \SP_{2T}(-C,-\eta)=-\SP_{2T}(C,\eta),
  \label{eq:Snegfmirror}
\end{equation}
which requires $\alpha_{2T}=c_{2T}=0$. Under this constraint, changing $\eta$
at fixed $C$ reverses $b_{2T}\eta$. Reversal of the total polarization is a
stronger condition: the phonon-odd contribution must exceed the $C$-odd
contribution in magnitude.

\subsection*{NEGF results}

The angular-momentum scan tests whether the prescribed coupling produces
spin polarization. For $L_0=0$, the mean-field spin--phonon term vanishes,
and the reference calculation with spin-independent contacts gives
$\SP_{2T}=0$. At the nonzero $L_0$ values in
Fig.~\ref{fig:S7}, the extracted $b_{2T}\eta$ is nonzero.

The contact comparison then tests the symmetry restriction. The configuration
that preserves Eq.~\eqref{eq:Snegfmirror} gives $\alpha_{2T}\approx0$ and
$c_{2T}\approx0$, whereas the right-contact modification in panel (d) produces
nonzero constant and mixed coefficients when the joint mapping is broken.
These comparisons concern the dependence on angular-momentum and chirality
reversal, not numerical equality with the FSSH coefficients. The two methods
use different phonon descriptions, transport observables, and values of
$\gsr$.

\bibliography{references}